\documentstyle[prl,aps,multicol,graphics]{revtex}
\begin{document}

\draft
\flushbottom

\title{Intensity Distribution of Modes
in  Surface Corrugated Waveguides}

\author{A. Garc\'{\i}a-Mart\'{\i}n, J.A. Torres,
J.J. S\'{a}enz}

\address{Departamento de F\'{\i}sica de la Materia Condensada
and Instituto de Ciencia de Materiales ``Nicol\'{a}s Cabrera'',
Universidad Aut\'{o}noma de Madrid, E-28049 Madrid, Spain.}

\author{M. Nieto-Vesperinas}

\address{Instituto de Ciencia de Materiales de Madrid.  Consejo Superior
de Investigaciones Cient\'{\i}ficas, Campus de Cantoblanco, E-28049 Madrid,
Spain. }

\date{11 November 1997}
\maketitle
\begin{center} {\bf Phys. Rev. Lett. 80, 4165 (1998)}\\ \copyright 1998
The American Physical Society \end{center}

\begin{abstract}

Exact calculations of transmission and reflection coefficients in
surface randomly corrugated optical waveguides are presented. As the
length of the corrugated part of the waveguide increases, there is a
strong preference to forward coupling through the lowest mode.
An oscillating behavior of
the enhanced backscattering as a function of the wavelength is
predicted. Although the transport is strongly non isotropic,
the analysis of the probability distributions of the
transmitted waves confirms in this configuration
distributions predicted by Random Matrix
Theory for volume disorder.
\\

\end{abstract}

\begin{multicols}{2}
%
%

Defects and inhomogeneities in waveguides are a subject of increasing
research for its influence on the propagation of both classical and
quantum waves\cite{mogo1}.
Volume defects have recently been studied in connection with the
propagation of electrons through two-dimensional (2D)
wires\cite{mogo2,NieEPL,PelosPRB96}.
On the other hand, slight
surface roughness has been considered both in electron transport through
thin films \cite{McGurn} as well as in
optical fibers and
waveguides \cite{Marcuse}, mainly in connection with
attenuation.  Recently, some analogies between electron transport
and propagation of diffuse light through surface corrugated waveguides
have been put forward\cite{APL}.
However, the assumption of
diffuse incoming waves limits the study to those
systems in which the incident modes are mutually incoherent.

In this letter we  address
the more practical situation in light propagation where a single
incident mode couples with different forward and backward modes
\cite{Ladou}.
As we shall see, the
scattering from rough,
perfectly conducting, walls (see Fig.1)
induces a strong and rich coupling
to forward and backward modes.
The coupling characteristics present
significant differences with respect to the coupling induced by
volume defects.  For any incoming mode, and when the length of the
system is larger than the  localization length $\xi$, there is a
strong preference for the forward propagation through the lowest mode.
The coupling to backward modes presents a very interesting behavior,
depending on the incident mode.  The ``external'' modes (defined as
those propagating modes with either the smallest or the largest transversal 
momentum)
present enhanced backscattering factor larger than the others.
In
the case of
the lowest mode, this factor exhibits remarkable oscillations as a
function of the wavelength.
We shall also present
an extensive analysis of the
statistical properties of the different transmission coefficients. As
far as we know, this
has been carried out only for the case of volume defects. In the diffusive
regime, the theoretical predictions of Ref.
\cite{NieRo} (and subsequently, Ref. \cite{KoKa}) have
recently been confirmed experimentally \cite{Genack}. A crossover from
either the Rayleigh and Gaussian statistics in the diffusive regime
to lognormal statistics in the localized regime, expected from numerical
simulations \cite{Edrei},
has been analytically deduced
in Ref. \cite{Bee}. Although the
mode coupling is quite different in the case of surface roughness,
our calculated distributions of
transmitivitties are fully consistent with the predicted behavior for
bulk disordered waveguides.

The corrugated part of the waveguide, of total length
$L$ and perfectly reflecting walls, is composed of $n$ slices of length
$l$. The width of each
slice has random values uniformly distributed between $W_0-\delta$ and
$W_0+\delta$ about a mean value $W_0$.  We shall take $W_0/\delta= 7$
and $l/\delta = 3/2$.
The main transport properties do not depend on the particular choice of
these parameters, however.  We consider
s-polarized waves with the
electric field parallel to the surface grooves (TE modes).  Transmission and
reflection coefficients are exactly calculated by solving the 2D wave
equation by mode matching at each slice, together with a generalized
scattering matrix technique\cite{APL,Weiss,JosanMetodo}.

For a given incoming mode $i$,
the transmission ($T_{ij}$) and reflection ($R_{ij}$) coefficients are
defined by
\begin{eqnarray}
T_{ij}=\frac{\Phi_{j}^{Forw}}{\Phi_{i}^{in}}\ , \mbox{ and }
R_{ij}=\frac{\Phi_{j}^{Back}}{\Phi_{i}^{in}}\ ,
\end{eqnarray}
$\Phi_{j}^{Forw}$  being the total flux transmitted into the forward
outgoing mode $j$, $\Phi_{j}^{Back}$ the total flux reflected into the
backward outgoing mode $j$, and $\Phi_{i}^{in}$ the total flux of the
incident mode.  The total transmission for the incoming mode $i$, is
given by $T_i = \sum_{j} T_{ij}$, and $G = \sum_{i} T_i$  is the
normalized transmittance (conductance).
Ensemble averages, denoted by $\langle . \rangle$, are
performed over a thousand realizations of the corrugated waveguide
(unless otherwise stated).

%
%


The analysis of the {\em conductance} of surface
corrugated waveguides is qualitatively similar to the
electron transport in disordered nanowires.
For short
lengths
the transport regime is diffusive
with an effective
mean free path \cite{APL}
$\ell = (N \partial
\langle 1/G \rangle/\partial L)^{-1}$, where $N$ is the number of
propagating modes. The transport regime changes
as $L$ becomes of the order of the localization length
\cite{Localization}
$
\xi =
-\left( \partial \langle \ln (G) \rangle /
\partial L \right)^{-1},
$
with
$\xi = N\ell$ within the numerical accuracy \cite{APL}.
However,
the analysis of the transmission coefficients $T_{ij}$ reveals
qualitative differences  between random surface and volume
scattering.

As a typical example, let us consider that the ratio between the waveguide 
width and  the wavelength is $W_0/\lambda=2.6$. This allows five propagating
modes. In Fig. 1 we plot
$\langle \ln(T_{ij}) \rangle$,
as a function of the normalized length $L/W_0$
for two  different incident modes $i=1$ and $i=3$
(Fig. 1 shows the average over
100 different
realizations for each length $L$).
All curves present the same linear dependence at large $L$.
The  inverse of the slope defines a  length $\xi_{ij} = -
\left( \partial \langle \ln(T_{ij}) \rangle / \partial L \right)^{-1}$
which {\em does not depend
on the incoming or outgoing modes} ($\xi_{ij}/W_0 = 34.4 \pm 0.7$ for
the structure calculated in Fig. 1).
Exactly the same dependence is found for $\langle \ln T_i \rangle$.
As a general result, and
within the numerical accuracy, $\xi_{ij}=\xi_i=\xi=N\ell$.

In contrast with the expected isotropic mode distribution induced by
scattering from volume defects, the calculation of
$T_{ij}$ and $T_i$ shows a clear non isotropic distribution, even in the
diffusive
regime ($\ell \le L \le \xi$) as shown in Fig. 2(a-c). In the localized
regime ($L > \xi$), for any incoming mode $i$ there is
a strong preference to couple with the lowest transmitted mode $j=1$.
The coupling to forward modes  decreases as their transversal
(longitudinal) momenta increase (decrease).

%
%

Concerning
backward modes,
Fig. 1 shows
the averaged reflection coefficients
$\langle R_{ij} \rangle$
as  functions of the length $L$.
Enhanced backscattering
effects (EB) can be clearly seen in this figure.
Coherent EB arising in the light scattering both from
dense media
\cite{ecb} and rough surfaces
\cite{ecbs} has been intensively investigated
\cite{mogo1}. For a waveguide with $N$ propagating modes,
Random Matrix Theory (RMT) predicts
\cite{Mello1,Beenakker-rev} that
the EB peak is 2 the scattering to any
other channel, irrespective of the incoming mode.
However, although not explicitly discussed,
numerical calculations in waveguides with volume disorder
\cite{PelosPRB96} show factors other than 2 for EB of
``external'' modes.
In order to quantify the EB peak,
we define, for
a given incoming mode $i$,
an {\em enhanced backscattering factor} $\eta_i$ as,
\begin{eqnarray}
\eta_{i}= (N-1) \frac{\langle R_{ii}\rangle}{\left(\sum_{j\neq i}\langle
R_{ij}\rangle\right)}\ .
\end{eqnarray}
Fig. 2d shows $\eta_i$ versus $L$.
While the ``central'' modes ($i=2,3,4$) present a factor $\eta \approx
2$, the ``external'' ($i=1,5$) modes have a much larger enhanced
backscattering factor ($\approx 4$).
This qualitative behavior, which also confirms indications for
electron propagation in disordered wires \cite{PelosPRB96}, depends on
the wavelength.
The backscattering factor for the lowest mode $\eta_1$ oscillates above
$\eta \approx 2$ as the
wavelength $\lambda$ decreases (i.e. $W_0/\lambda$ increases) having its
maxima close to half integers of $W_0/\lambda$, which correspond to the
appearance of a new propagating mode in the waveguide\cite{tobe}.
This behavior is highly correlated
with the oscillations both in the mean free path $\ell$ and in the
localization length
$\xi$ \cite{APL} (the maxima in $\eta$ correspond to minima in $\xi$).
Just after the appearance of a new mode  the
backscattering dominates at this
new appearing channel. This gives a very large $\eta$ factor at the
onset of the new mode, which slowly decreases to $\eta =2$ as $\lambda$
decreases. By contrast, the ``central'' modes present an almost constant
enhanced backscattering factor near 2.

Let us now discuss the intensity probability distributions of 
the transmitted waves.
As shown above, the behavior of the
different averaged transmitivities for a surface disordered waveguide
is qualitatively different from that obtained for volume
disordered wires. Therefore, the question now is: does surface disorder
change the
probability distributions from those expected in bulk disordered systems?.
In the case of random wires,
diagrammatic techniques combined with RMT
\cite{NieRo,KoKa} show that in the diffusive regime the
speckle pattern probability density $P(T_{ij})$ follows a Rayleigh
statistics with stretched exponential tails, while
the distribution $P(T_i)$ is Gaussian with tail deviations.
Nonperturbative calculations done in absence of time
reversal symmetry predict  that these distributions
evolve into the same lognormal
distribution as the length $L$ increases beyond the localization length
$\xi$ as expected from the numerical simulations of Ref. \cite{Edrei}.

Fig. 3a contains the distribution
$P(T_i/\langle T_i \rangle)$ for different values of
$\langle T_i \rangle$.
Although for a fixed length, the
distributions vary from mode to mode, we find that,
as long as
$L \le \xi$, the distribution
does not depend on the incoming mode $i$ neither on the length $L$,
but only depends on the
averaged transmission coefficients.
The inset in Fig. 3a shows
$P(T_i/\langle T_i \rangle)$
for $i=3,4,5$
all having the same average,
$\langle T_i \rangle=0.16$,
but corresponding to different lengths of the disordered region
(see Fig. 2c).
For the same average,
modes $i=1,2$ are localized and have a different distribution (see
below).
The  speckle distribution
$P(T_{ij}/\langle T_{ij} \rangle)$, in the diffusive regime follows an
almost perfect Rayleigh statistics $P(x) = \exp(-x)$
(see Fig. 3b). Clear deviations from this exponential distribution are
observed for $L \le \ell$ (not shown in the figure) and for $L \ge \xi$.
An interesting point is that the dependence of $\langle T_{ij} \rangle$
with length does not fix, by itself, the transport regime. While for
$\ell < L < \xi$ all coefficients follow Rayleigh statistics, their 
dependence with $L$ change completely from one to another 
as can be observed in Fig. 2a,b.

Except for either small lengths $L \le \ell $
or  large transmitivities
($\langle T_i \rangle \ge 1/2$),
the qualitative behavior of  both
$P(T_i/\langle T_i \rangle)$ and
$P(T_{ij}/\langle T_{ij} \rangle)$
is basically the same as that obtained for
disordered wires \cite{NieRo,KoKa,Bee}
(compare our Fig. 3 with Fig. 1 of reference \cite{Bee}).
Moreover, both distributions evolve
into the  same lognormal
distribution as the length $L$ increases beyond the localization length.
This is shown in Fig. 4 where
$P(\ln({\large \tau}))$ is plotted versus $\ln({\large \tau})$, 
${\large \tau} = T_i , T_{ij}$,
for different values of $\langle \ln({\large \tau}) \rangle$.
Solid lines fitted there represent the lognormal distribution
\begin{equation}
P(x) = \left(\frac{1}{ \pi 2 \sigma_{\langle x \rangle}^{2}}
\right)^{1/2} \exp \left[ -
\frac{ (x-
\langle x \rangle)^2}{2 \sigma_{\langle x \rangle}^{2}}
\right]
\end{equation}
with $x=\ln(T_i), \ln(T_{ij})$.
and var$(x) = \sigma_{\langle x \rangle}^{2} = (3/2) |\langle x
\rangle|$.
In the strong localization limit
$(L \gg \xi)$, where
$|\langle \ln({\large \tau}) \rangle| \approx L/\xi =
L/(N\ell)$, we obtain  $\sigma_{\langle x \rangle}^{2} \approx (3/2) L/\xi$.
Nevertheless,
although the dependence is the same as expected from RMT
\cite{Bee}, the variance of the calculated distribution is  3/2
the mean in contrast with the expected value of twice the mean (cf.
(11) of reference \cite{Bee}).
However, the analysis of the conductance $G$ in the localized regime
shows a
lognormal distribution
with a mean which is half
the variance, in perfect agreement with RMT predictions
\cite{Imry,Beenakker-rev}. This is shown in
the inset of Fig. 4.
The solid lines show the corresponding
lognormal distribution (eq. 3 with $x=\ln(G)$) with
var$(x) = \sigma_{\langle x \rangle}^{2} =  2    |\langle x
\rangle|$.

In conclusion, we have analyzed the coupling to forward and backward
modes in surface corrugated waveguides.
The coupling with backward modes
yields  enhanced backscattering peaks, which can be larger than those
theoretically
predicted for the case of the lowest and highest incident modes.
We have shown important differences in the transport coefficients with
respect to the case of volume scattering. However, the probability
distributions are the same as those obtained
for bulk disordered waveguides and wires.
In particular we have found a crossover from Rayleigh and Gaussian
distributions in the diffusive regime to the same lognormal distribution
in the localized regime in good agreement with the predictions of Random
Matrix Theory.

We would like to thank E. Bascones, A.J.  Caama\~{n}o, G. G\'{o}mez-Santos and T.
L\'{o}pez-Ciudad for stimulating discussions.
This work has been supported by the Direcci\'{o}n General de Investigaci\'{o}n
Cient\'{\i}fica y T\'{e}cnica (DGICyT) through Grant No.  PB95-0061 and the
European Union. A. G-.M. 
acknowledges partial financial support from the postgraduate grant 
program of the Universidad Aut\'{o}noma de Madrid.

\begin{figure}


\narrowtext
\caption{
Averaged transmission $(T_{ij})$ and reflection $(R_{ij})$ coefficients
(incoming mode $i$; outgoing mode $j$)
as a function of the normalized length of the disordered region $L/W_0$.
Inset: Schematic view of the system under consideration.}
\label{Fig. 1}
\end{figure}


\begin{figure}


\narrowtext



\caption{
(a) Averaged transmission coefficients $\langle T_{ij}\rangle$ versus
$L/W_0$ (incoming mode $i=1$, outgoing mode $j$).
(b) The same as (a) for i=3.
(c) Averaged transmission coefficients $\langle T_{i}\rangle$  
versus $L/W_0$. At $\langle T_{i}\rangle=0.16$ (horizontal dashed line)
incoming modes  $i=3,4,5$ have the same probability densities as shown 
in the inset of Fig. 3.
(d) Enhanced backscattering factor $\eta_i$ versus
 of $L/W_0$.
 Vertical long-dashed and dotted-dashed lines are drawn at
$L=\ell$ and $L=\xi$ respectively.}  
\label{Fig. 2}
\end{figure}


\begin{figure}


\narrowtext



\caption{
(a) Distribution of $x=T_i/\langle T_i \rangle$ for different
values of $\langle T_i \rangle$.
The inset  shows the
distribution for $i=3,4,5$ having the same average $\langle T_i \rangle
\approx 0.16$ (see Fig. 2c).
The smoothed curves are superimposed to the histograms to guide the eye.
(b) Distribution of $x=T_{ij}/\langle T_{ij}
\rangle$ for different lengths $L/\xi$. The straight line corresponds to
the exponential distribution.}

\label{Fig. 3}
\end{figure}


\begin{figure}


\narrowtext



\caption{
Distribution of $\ln (\tau)$ with $\tau =T_i$, $T_{ij}$
for different values of $\langle \ln (\tau) \rangle$.
The inset  shows the
distribution of the conductance $G$ for two different lengths of the
disordered region. Continuous lines are fittings to lognormal
distributions having  $\sigma_{\langle x \rangle}^{2}=3/2 \langle x \rangle$, 
for $x=\ln(\tau)$, and $\sigma_{\langle x \rangle}^{2}=2 \langle x \rangle$ 
for $x=\ln(G)$.}
\label{Fig. 4}
\end{figure}
\end{multicols}

\end{document}